\documentclass[
  aip,
  author-year,
  showpacs,
  preprintnumbers,
  showkeys,
  amsmath,amssymb,
  tightenlines,
]{revtex4-1}

\usepackage[utf8]{inputenc}
\usepackage{bm}
\usepackage{datetime}
\usepackage{tabularx}
\usepackage[pdftex]{hyperref}
\usepackage[pdftex]{graphicx}

\newcommand\tabref[1]{\tablename~\ref{#1}}
\newcommand\secref[1]{sect.~\ref{#1}}
\newcommand\figref[1]{\figurename~\ref{#1}}

\begin{document}

\title{Compressive consolidation of strongly aggregated particle gels}
% running title : fractality loss through compression

%  \date{\shortdate\today \, \ampmtime }
\date{\today}
\author{Ryohei \surname{Seto}}
\email{setoryohei@me.com}
\affiliation{%
  Levich Institute, The City College of New York,
  140th Street and Convent Avenue, New York, NY 10031}
\affiliation{%
  Max Planck Institute for Polymer Research, Ackermannweg 10, 55128 Mainz, Germany}
\author{Robert \surname{Botet}}
\affiliation{%
  Laboratoire de Physique des Solides,
  UMR8502, Université Paris-Sud, Bât 510, 91405 Orsay, France}
\author{Martine \surname{Meireles}}
\affiliation{%
  Laboratoire de Génie Chimique, UMR 5503, Université Toulouse III, 31 062 Toulouse, France
}
\author{Günter K. \surname{Auernhammer}}
\affiliation{%
  Max Planck Institute for Polymer Research, Ackermannweg 10, 55128 Mainz, Germany}
\author{Bernard \surname{Cabane}}
\affiliation{%
  PMMH, UMR 7636, ESPCI, 75231 Paris cedex 05, France}

\begin{abstract}
The compressive yield stress of particle gels
shows a highly nonlinear dependence on the packing fraction.
We have studied continuous compression processes,
and discussed the packing fraction dependence
with the particle scale rearrangements.
The 2D simulation of uniaxial compression was applied to fractal networks,
and the required compressive stresses
were evaluated for a wide range of packing fractions
that approached close packing.
The compression acts to reduce 
the size of the characteristic structural entities 
(i.e. the correlation length of the structure).
We observed three stages of compression:
(I) elastic-dominant regime;
(II) single-mode plastic regime,
where the network strengths  are 
determined by the typical length scale and the rolling mode;
and (III) multi-mode plastic regime,
where sliding mode and connection breaks are important.
We also investigated the way of losing the fractal correlation under compression.
It turns out that both fractal dimension $D_{\mathrm{f}}$ 
and correlation length $\xi$
start to change from the early stage of compression,
which is different from the usual assumption in theoretical models.
\end{abstract}

\maketitle

\section{Introduction}

Particle gels form in suspensions
if short-range attractions act between contacting surfaces of particles.
Depending on the strength of the attraction,
one can distinguish them as two types:
strongly and weakly aggregated particle gels~\citep{Larson_1999}.
In strongly aggregated particle gels, on which we focus in this work,
particles once joined hardly separate at room temperature~\citep{Witten_2004}.
The attractions also prevent tangential displacements
of contacting particles~\citep{Dinsmore_2010},
thus particles form sample-spanning loose networks.
Their remarkable and complex mechanical properties 
under external stress are essentially due to 
the response of a deformable disordered network formed with all the particles.
%

%
%%%%%%%%%%%%%%%%%%%%%%%%%%%%%%%%%%%%%%%%%%%%%%%%%%%%%%%%%%%%%%%%%%%%%%%%%%%%%%%%%%%%%%%%%%%%%%%%%%%%

Under shear stress, 
one can sketch out deformation of a particle gel as follows~\citep{Nguyen_1992}:
\begin{itemize}
\setlength{\parskip}{-5pt}
  \item 
  the system is elastic under very low stresses. 
  If the stress is released, the system comes back to the original shape;
  \item 
  for larger stress, a particle gel may undergo finite deformations, 
  though it does not flow. 
  This is the plastic regime, and it gives malleability;
  i.e., the system mostly keeps the deformed shape if stress is released;
  \item 
  beyond a yield stress, the particles network is ruptured and 
  the particle gel flows.
\end{itemize}
Fluids exhibiting these behavior 
are called \emph{yield stress} (or \emph{Bingham plastic}) fluids~\citep{Nguyen_1992,Denn_2011}.
Since the yield stress is the highest stress that can prevent flow,
it can be considered as the strength of particle gels.
Naturally, this strength relates to the strength of the particle connections.
In a particle gel under shear stress, 
transmitting stresses break a portion of particle connections
and the sample spanning network is torn apart.
The yield stress of particle gels is higher 
when the packing fraction of particles is larger.
In many cases, this variation appears to follow
a power law with exponents between 4 and 5~\citep{Channell_1997}.

%%%%%%%%%%%%%%%%%%%%%%%%%%%%%%%%%%%%%%%%%%%%%%%%%%%%%%%%%%%%%%%%%%%%%%%%%%%%%%%%%%%%%%%%%%%%%%%%%%%%

Under compressive stress,
the behavior of a particle gel looks different, 
due to confinement~\citep{Kretser_2003,Stickland_2009}.
Compressive stress appears naturally in a number of applications,
for example paste drying~\citep{Brown_2002} or ceramic processing~\citep{Lewis_2000}.
Osmotic compression is a generic important case.
In this kind of compression process,
liquid removal tends to pack the particles into a smaller volume.
As long as the applied stress is 
lower than the strength of the gel,
it will retain the volume.
If the stress exceeds a compressive yield stress $P_{\mathrm{y}}$,
network restructuring leads to the loss of a part of the liquid phase,
and particles reorganize into a smaller volume~\citep{Buscall_1987a}.
The deformation stops when the mechanical resistance
of the network equals the stress.
Thus, unlike the shear case, 
deformation does not continue 
with a constant applied stress~\citep{Buscall_1987}.
The growth of mechanical resistance, i.e. compressive consolidation,
also shows a highly nonlinear dependence on the packing fraction,
typically power-law relations $P_{\mathrm{y}}\sim \phi^{\lambda}$
with exponents between 3.2 and 4.5%
~\citep{Miller_1996,Green_1997,Gisler_1999,Madeline_2007,Parneix_2009}.

%%%%%%%%%%%%%%%%%%%%%%%%%%%%%%%%%%%%%%%%%%%%%%%%%%%%%%%%%%%%%%%%%%%%%%%%%%%%%%%%%%%%%%%%%%%%%%%%%%%%

%
The highly nonlinear dependency of the compressive yield stress $P_{\mathrm{y}}$
on the packing fraction $\phi$
relates to the heterogeneity of the network structure of particles~\citep{Buscall_1987a}.
The fractal model~\citep{Brown_1986,Shih_1990a,Potanin_1996,Mewis_2011}
successfully captures
such packing-fraction dependence of rheological proprieties
by considering the characteristic structure of 
strongly aggregated particle gels.
In the model, a particle gel is assumed to be an assembly of fractal flocs\textemdash
such structural correlation 
is evidenced by scattering observations~\citep{Dietler_1986,Poon_1997}. %Schaefer_1984
The entire sample is filled up with
flocs of a fractal dimension $D_{\mathrm{f}}$,
and the correlation length $\xi$ is equivalent or proportional 
to the average size of the flocs.
Thus, the correlation length $\xi$ can be associated 
with the packing fraction $\phi$
by using the fractal dimension $D_{\mathrm{f}}$:
$\xi/a = c \phi^{-1/(d-D_{\mathrm{f}})}$,
where $d$ and $c$
are the spatial dimension and proportional coefficient.
Under this assumption,
a gel of a higher packing fraction
consists of a higher number of smaller flocs
of the same fractal dimension $D_{\mathrm{f}}$.
By deducing average mechanical responses of the flocs,
the fractal model predicts a power-law dependence 
of the shear modulus on $\phi$
with an exponent $(d+D_{\mathrm{bb}})/(d-D_{\mathrm{f}})$,
where $D_{\mathrm{bb}}$ is the fractal dimension of the backbone structure.
Although this theoretical prediction 
has often been cited in the literature, 
we should pay attention to the scope of the model.
Since fractal correlation yields in the formation process of particle gels,
there is no clear reason for $D_{\mathrm{f}}$ and $D_{\mathrm{bb}}$ 
to be kept constant during the compression process.
Actually, some scattering observations indicate
that the compression process changes the local structure 
(fractal dimension)~\citep{Madeline_2007,Parneix_2009}.
To make progress in macroscopic modeling of aggregated 
colloidal or particle gels~\citep{Flatt_2006,Lester_2010},
we need to settle such microstructural issues.
%

%%%%%%%%%%%%%%%%%%%%%%%%%%%%%%%%%%%%%%%%%%%%%%%%%%%%%%%%%%%%%%%%%%%%%%%%%%%%%%%%%%%%%%%%%%%%%%%%%%%%

There is a recent two-dimensional(2D) simulation work 
to support the application of the fractal model 
to compression processes~\citep{Gilabert_2008}.
Gels which were 
(i) directly prepared to the packing fraction $\phi$,
or (ii) prepared by compression
from a lower packing-fraction gel $\phi' <\phi$
showed the same structural characteristics. 
However, the investigated range of packing fractions
was relatively high ($0.47 < \phi < 0.78$).
Since the fractal model targets on strongly aggregated gels,
we should compare simulations at lower packing fractions.
Indeed, our 2D simulation investigates 
a wider range of packing fractions ($0.12 < \phi < 0.74$).
%

%%%%%%%%%%%%%%%%%%%%%%%%%%%%%%%%%%%%%%%%%%%%%%%%%%%%%%%%%%%%%%%%%%%%%%%%%%%%%%%%%%%%%%%%%%%%%%%%%%%%

This article reports results of numerical simulation
for compression of particulate networks
to discuss the consolidation behavior with restructuring.
Experimentally observed time evolution of compressed gels
should be affected by both the hydrodynamic and 
network resistances~\citep{Manley_2005a,Kim_2007,Brambilla_2011}.
%Lester_2005
But, we here focus on studying the latter;
our simulation work investigates 
the quasi-static structural evolution 
determined by the balance
between external stresses and network strength at long time.
This limitation to quasi-static aspects is common with 
the above-mentioned fractal model.
So, compression considered in this work is sufficiently slow,
such as osmotic compression.
A contact model is employed to simulate
sticky and breakable connections between contacting particles.
In granular physics,
there are many prior studies
to investigate cohesive granular system%
~\citep{Iwashita_1998,Tomas_2007}.
Their simulation methods are an extension of 
the discrete element method~\citep{Cundall_1979};
the rolling friction has been 
additionally included into the contact model.
Though our target is a suspension system,
the spirit of the modeling is almost the same.
In granular physics,
there are also some studies on compression of cohesive powders%
~\citep{Kadau_2002,Bartels_2005,Wolf_2005,Gilabert_2007,Gilabert_2008},
which obtained many common insights with our study.
%
%But this work is unique in its focus on restructuring 
%of networks possessing long correlation lengths.
%
2D simulation results are reported here.
One obvious benefit of studying 2D systems 
is easy recognition of the structure changes.
Though the density-density correlation is an effective tool to capture
special sorts of heterogeneous structures,
it might not capture all features of random structures. 
2D experimental system has also been studied
to observe the microstructure change under shear
~\citep{Masschaele_2009}.
Though we do not expect qualitative difference between 2D and 3D systems,
this point should be checked by 3D simulation of the same modeling,
which we hope to report in future work.
%

%2D systems are not limited as subjects of simulation studies,
%but there are some experimental investigations%
%~\citep{Masschaele_2009,Cicuta_2009,Berhanu_2010,Park_2011}.
%
%

\section{Method}

\subsection{Initial configuration: fractal gel}
\label{sec_preparation_fractal_gel}

We use fractal aggregation models to generate 
the initial configurations.
Networks of particles are formed in 
a rectangle box of sides $L_x$ and $L_y$ with periodic boundary conditions.
We combine two processes.
The first step models a formation process of 
fractal clusters in the dilute limit;
the reaction-limited hierarchical cluster-cluster aggregation model
was employed~\citep{Jullien_1987}.
Each cluster consists of $N_{\mathrm{floc}}$ particles.
The second step is to build a network with them.
$N_{\mathrm{cl}}$ of the generated clusters are randomly placed 
in the box while avoiding overlap with each other, 
and then they are moved with a random walk.
Collisions of clusters always lead to coalescence,
so a single cluster is formed 
in the periodic boundary space after a while
(see, e.g., \figref{fig_snapshots}~(a), below).
Thus, we can generate
random networks of a packing fraction 
$\phi = \pi a^2 N_{\mathrm{cl}} N_{\mathrm{floc}}  / L_x L_y $,
where $a$ is the radius of a particle.

%%%%%%%%%%%%%%%%%%%%%%%%%%%%%%%%%%%%%%%%%%%%%%%%%%%%%%%%%%%%%%%%%%%%%%%%%%%%%%%%%%%%%%%%%%%%%%%%%%%%

\begin{figure}[htb]
  \centering
  \includegraphics[height=156pt]{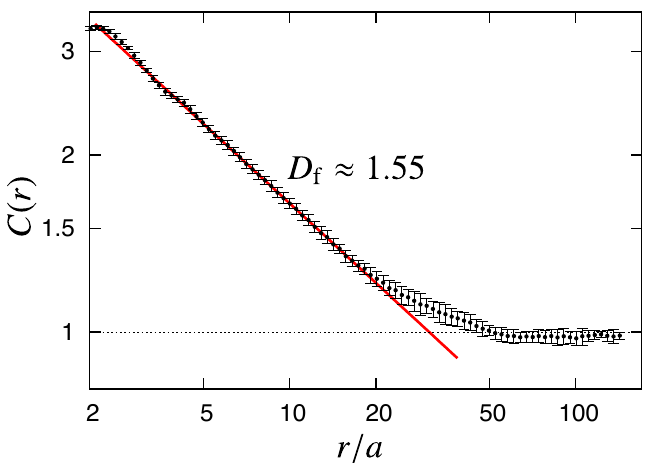}
  \caption{
    The density-density correlation functions
    were evaluated with 
    the fractal networks of $\phi \approx 0.12$ 
    prepared in the box $(L_x, L_y) = (300a, 500a)$.
    The averages and standard deviations were taken over 10 samples.
    The red line indicates the slope corresponding
    to $D_{\mathrm{f}}\approx 1.55$.
  }
  \label{223812_16Oct12}
\end{figure}

One can deduce structural features of the formed networks
by evaluating the density-density correlation function $C(r)$ 
(See appendix \ref{appendix_density-correlation}).
We prepared 10 samples of $N=5760$ particles
in a box $(L_x, L_y) = (300 a, 500 a)$.
The packing fraction is $\phi \approx 0.12$.
\figref{223812_16Oct12} shows
the averages and standard deviations 
of their correlation functions $C(r)$.
A power-law decay of correlation is seen in the finite range $ 6 < r/a <  15$.
The slope indicates fractal dimension $D_{\mathrm{f}} \approx 1.55$,
which is almost the same fractal dimension as 
the clusters generated by the aggregation model.
Though a crossover behavior is seen in the range $15 < r/a < 70$,
the correlation decays out after that.

\subsection{Contact forces}
\label{112400_22Dec12}

Our contact model is designed to simulate strongly aggregated particle gels.
The contact model used in the original discrete element method (DEM)~\citep{Cundall_1979}
takes into account friction forces for the sliding mode,
which is essential to simulate dense granular systems.
For cohesive granular systems, the rolling friction has been added%
~\citep{Iwashita_1998}.
This extension is also required 
to reproduce the behavior of particle gels having chain-like local structures.
Our implementation is similar to the model used for cohesive granular systems.
Since the dynamic friction, which is modeled in typical contact models,
is unclear for colloidal systems, 
this part is different in our model.
Though a brief summary will be given below,
the detailed descriptions of our implementation
were given in another paper~\citep{Seto_2012a}.
%

%%%%%%%%%%%%%%%%%%%%%%%%%%%%%%%%%%%%%%%%%%%%%%%%%%%%%%%%%%%%%%%%%%%%%%%%%%%%%%%%%%%%%%%%%%%%%%%%%%%%

%
We assume that interactions acting among particles
are only short range forces.
No interparticle interaction acts between two particles
before getting into contact.
Once the gap between two particles becomes zero, 
a cohesive force starts to act between them.
For simplicity, our simulation model assumes immediate bond generation,
which is expected to give the upper limit of the consolidation behavior.
When compression-rate dependence is problem,
the condition of bond generation can be crucial.
Finite compression rates can suppress bond generation, 
leading to a weaker consolidation.

%

%%%%%%%%%%%%%%%%%%%%%%%%%%%%%%%%%%%%%%%%%%%%%%%%%%%%%%%%%%%%%%%%%%%%%%%%%%%%%%%%%%%%%%%%%%%%%%%%%%%%

The forces and moments are assumed to be Hookian:
Let us consider the contact interactions between particles $i$ and $j$.
The center-to-center distance between them is 
$r^{(i,j)} \equiv |\bm{r}^{j}-\bm{r}^{i}|$,
and the gap $\delta^{(i,j)} \equiv (r^{(i,j)}-2a)$.
The normal (separation/overlap) displacement 
is related to the normal element of the contact force,
\begin{align}
  \bm{F}_{\mathrm{N}}^{(i,j)}
  = 
  \big\{k_{\mathrm{N}} \delta^{(i,j)}
    + C \vartheta(-\delta^{(i,j)}) \bigl(\delta^{(i,j)}\bigr)^3  \big\} \bm{n}^{(i,j)},
  \label{eq_f_normal_foo}
\end{align}
where $\vartheta(x)=1$ for $x > 0$ and $\vartheta(x)=0$ otherwise.
$\bm{n}^{(i,j)}$ is the normal unit vector defined by
$\bm{n}^{(i,j)} \equiv (\bm{r}^{(j)}-\bm{r}^{(i)})/r^{(i,j)}$.
The second term of eq.~\eqref{eq_f_normal_foo} is added 
to avoid significant overlap between particles.
The sliding displacement $\bm{d}^{(i,j)}$
relates to the tangential part of the contact force,
\begin{equation}
  \bm{F}_{\mathrm{T}}^{(i,j)}
  =
  k_{\mathrm{S}}\bm{d}^{(i,j)}.
\end{equation}
The sliding displacement $\bm{d}^{(i,j)}$
is the tangential projection 
of the deviation vector between the original contact points
$\bm{r}^{(i)} + a \bm{\xi}^{(i,j)}$
and 
$\bm{r}^{(j)} + a \bm{\xi}^{(j,i)}$,
\begin{equation}
  \bm{d}^{(i,j)} \equiv
  a \Bigl\{
    (\bm{\xi}^{(j,i)}-\bm{\xi}^{(i,j)})
    - 
    (\bm{\xi}^{(j,i)}-\bm{\xi}^{(i,j)}) 
    \cdot \bm{n}^{(i,j)} \bm{n}^{(i,j)}
    \Bigr\},
\end{equation}
where the vectors $\bm{\xi}^{(i,j)}$ and $\bm{\xi}^{(j,i)}$
indicate the original contact points from the respective centers 
of the particles and are fixed with them. 
The bending angle
$\varphi^{(i,j)} $
($\equiv - \sin^{-1} (\bm{\xi}^{(i,j)} \times \bm{\xi}^{(j,i)})_z)$
is related to the moment acting on the rolling mode,
\begin{equation}
  M_{\mathrm{R}}^{(i,j)}
  = k_{\mathrm{R}} a^2 \varphi^{(i,j)}.
\end{equation}

These elastic relations are broken
when the force or moment exceed threshold values.
We assume simple criteria given with three threshold values:
the critical normal force $F_{\mathrm{Nc}}$, 
critical sliding force $F_{\mathrm{Sc}}$, 
and critical rolling moment $M_{\mathrm{Rc}}$.
When the pulling force exceeds the value;
$\bm{F}_{\mathrm{N}}^{(i,j)}\cdot \bm{n}^{(i,j)} > F_{\mathrm{Nc}}$,
the bond is broken and the two particles separate.
When the tangential force or rolling torque exceeds 
the threshold value;
$|\bm{F}_{\mathrm{T}}| > F_{\mathrm{Sc}}$ or 
$|M_{\mathrm{R}}| > M_{\mathrm{Rc}}$,
both the sliding displacement $\bm{d}$ and 
rolling angle $\varphi^{(i,j)}$ are reset to zero,
which releases the tangential parts of the stored energy.
The maximum strains are related to the threshold 
forces and moment as follows:
$\delta_{\mathrm{c}} \equiv F_{\mathrm{N}} / k_{\mathrm{N}}$, 
$ d_{\mathrm{c}} \equiv F_{\mathrm{S}} / k_{\mathrm{S}}$, 
and $\varphi_{\mathrm{c}} \equiv  M_{\mathrm{Rc}} / (k_{\mathrm{R}} a^2)$.
%
%We will test this model
%by the above-mentioned three-point bending geometry in section \ref{sec_three-point_bending}.
%

%%%%%%%%%%%%%%%%%%%%%%%%%%%%%%%%%%%%%%%%%%%%%%%%%%%%%%%%%%%%%%%%%%%%%%%%%%%%%%%%%%%%%%%%%%%%%%%%%%%%

It must be noted that in our simple model
the bond strength of the tangential mode does not depend on 
the normal load.
However, 
the enlarged contact area due to the normal load
may change the tangential strengths~\citep{Dominik_1995}.
Our neglect of this effect can affect simulations at higher packing fractions.

\subsection{Equation of motion}

A particle $i$ may be in contact with other particles $j$.
The total contact force and torque acting on the particle $i$ 
can be written as follows:
\begin{equation}
  \begin{split}
    \bm{F}_{\mathrm{C}}^{(i)}&= 
    \sum_{j} 
    \left(
      \bm{F}_{\mathrm{N}}^{(i,j)}
      + 
      \bm{F}_{\mathrm{T}}^{(i,j)}
    \right), \\
    %%%%%%%%%%%%%%%%%%%%%%%%%%%%%%%%%%%%%%%%%%%%%%%%%%%%%%%%%%%%%%%%%%%%%%%%%%%%%%%%%%
    T^{(i)}_{\mathrm{C}} &= 
    \sum_{j} 
    \left(
      M_{\mathrm{R}}^{(i,j)}
      +
      a\bm{n}^{(i,j)} \times \bm{F}_{\mathrm{T}}^{(i,j)}
    \right)
    .
  \end{split}\label{141319_10Nov12}
\end{equation}
%

%%%%%%%%%%%%%%%%%%%%%%%%%%%%%%%%%%%%%%%%%%%%%%%%%%%%%%%%%%%%%%%%%%%%%%%%%%%%%%%%%%%%%%%%%%%%%%%%%%%%
Particles are immersed in a viscous fluid,
so that hydrodynamic interactions also act on them.
For micro or nano-sized particles,
the particle Reynolds number is extremely small,
so the Stokes equations should be used 
to evaluate the hydrodynamic interactions among many particles~\citep{Kim_1991}.
However, we do not go into the detailed hydrodynamics here.
%qs
Hydrodynamic interactions in the Stokes regime
are scaled with the velocities of particles.
If the slow compression limit is considered,
flows induced by particle movements are negligible
in comparison with the contact forces.
They just work as energy dissipators.
Indeed, some experimental observations suggest that 
a solvent flow does not contribute the rearrangements of particles
in a slow sedimentation of colloidal gels~\citep{Brambilla_2011}.

%Due to this simplification,
%we cannot discuss about
%the compression-rate dependence 
%due to hydrodynamic resistance~\citep{Lester_2005}.

%%%%%%%%%%%%%%%%%%%%%%%%%%%%%%%%%%%%%%%%%%%%%%%%%%%%%%%%%%%%%%%%%%%%%%%%%%%%%%%%%%%%%%%%%%%%%%%%%%%%
As an approximation, the Stokes formula is used in this work;
the force and torque acting on a particle $i$
are proportional to 
the translational velocity $\bm{v}^{(i)}$
and angular velocity $\omega^{(i)}$, respectively:
\begin{equation}
  \begin{split}
    \bm{F}_{\mathrm{H}}^{(i)} &= - 6 \pi \eta_0 a \bm{v}^{(i)}, \\
    %%%%%%%%%%%%%%%%%%%%%%%%%%%%%%%%%%%%%%%%%%%%%%%%%%%%%%%%%%%%%%%%%%%%%
    T_{\mathrm{H}}^{(i)} &= - 8 \pi \eta_0 a^3 \omega^{(i)},
  \end{split}\label{141331_10Nov12}
\end{equation}
where $\eta_0$ is the viscosity of the solvent.
%

%%%%%%%%%%%%%%%%%%%%%%%%%%%%%%%%%%%%%%%%%%%%%%%%%%%%%%%%%%%%%%%%%%%%%%%%%%%%%%%%%%%%%%%%%%%%%%%%%%%%
Since the interparticle interactions
are assumed to be sufficiently stronger than thermal agitations,
Brownian forces are also neglected in this work.
Consequently, 
Newton's equations of motions are given as 
the sum of the contact forces \eqref{141319_10Nov12}
and hydrodynamic interactions \eqref{141331_10Nov12},
\begin{equation}
  \begin{split}
    & m\frac{\mathrm{d} \bm{v}^{(i)}}{\mathrm{d}t}
    = \bm{F}^{(i)}_{\mathrm{C}} + \bm{F}^{(i)}_{\mathrm{H}},  \\
    %%%%%%%%%%%%%%%%%%%%%%%%%%%%%%%%%%%%
    & \frac{2}{5}m a^2\frac{\mathrm{d} \omega^{(i)}}{\mathrm{d}t}
    = T^{(i)}_{\mathrm{C}} + T^{(i)}_{\mathrm{H}}.
  \end{split}\label{180838_18Nov12}
\end{equation}
Since particle inertia is very small for colloidal systems,
it is a good approximation 
to neglect the inertia terms (the left hand sides)
and solve the force balance equations~\citep{Ball_1997}.
As explained later (\secref{141932_2Nov12}),
however, we took a different strategy to find the static equilibriums.

\subsection{Uni-axial compression and static equilibrium}
\label{sec_uniaxial_compression}

The compression is externally controlled.
For simplicity, 
we focus on examining the uni-axial compression
along the $y$ axis in this work.
The periodic boundaries at $y=0$ and $y=L_y$
of the simulation box introduced in section~\ref{sec_preparation_fractal_gel}
are replaced by two flat walls.
These walls can move along the $y$-axis like pistons,
and they are assumed to be semipermeable;
liquid passes through them, while particles do not.
When particles get in contact with the walls,
they are rigidly fixed there;
particles contacting with the wall move together as one object.
The periodic boundaries along the $x$-axis are still used 
during the compression simulation.

%%%%%%%%%%%%%%%%%%%%%%%%%%%%%%%%%%%%%%%%%%%%%%%%%%%%%%%%%%%%%%%%%%%%%%%%%%%%%%%%%%%%%%%%%%%%%%%%%%%%

The stress-controlled compression, 
in which an external stress is applied to cause the compression,
is similar to typical experimental situations, such as osmotic compression.
However, due to the higher efficiency of the simulation, 
we use strain-controlled compression. 
This also allows us to have better statistics in our data. 
This work investigates the slow compression limit;
the time scale of compression is much longer than the relaxation time 
of particle motions.
In this case, the results obtained with the strain-controlled compression
are essentially equivalent to the ones with the stress-controlled compression.
%

%%%%%%%%%%%%%%%%%%%%%%%%%%%%%%%%%%%%%%%%%%%%%%%%%%%%%%%%%%%%%%%%%%%%%%%%%%%%%%%%%%%%%%%%%%%%%%%%%%%%

The compressive stress balancing the network resistance
is evaluated from the total forces acting on the walls from particles.
Since particles directly connecting to the walls (called wall particles)
no longer have any mechanical freedom,
the forces between wall particles $i$ and
particles $j$ contacting with them are summed up as follows:
\begin{equation}
  F_{\mathrm{wall}}
  = 
  \sum_{i}\sum_{j} \bm{F}_{\mathrm{C}}^{(i,j)}\cdot \bm{e}_y.
\end{equation}
When we refer to ``compressive stress'' in this 2D simulation work,
our simulation box is assumed to have a thickness of particle diameter $2a$.
Therefore, the compressive stresses are given as
$P_{\mathrm{top}} =  F_{\mathrm{wall}} / (2a L_x)$
and 
$P_{\mathrm{bot}} = - F_{\mathrm{wall}} / (2a L_x)$.
%

%%%%%%%%%%%%%%%%%%%%%%%%%%%%%%%%%%%%%%%%%%%%%%%%%%%%%%%%%%%%%%%%%%%%%%%%%%%%%%%%%%%%%%%%%%%%%%%%%%%%

Our compression simulations are performed in a stepwise manner
in order to obtain numbers of equilibriums by a single simulation run.
Compression steps are given by two increment ratios $w_0$ and $w$;
the initial compression from $\phi_0 $ to $ \phi_1$ 
is given as $\phi_1 = w_0 \phi_0$,
and the $k$-th packing fraction as $\phi_k = w^{k-1} \phi_1$.
The increment ratio $w$ is determined 
with the number of steps $k'$ from $\phi_1$ to $\phi_{\mathrm{max}}$:
$w = (\phi_{\mathrm{max}}/\phi_{1})^{1/(k'-1)} $.
The strain-controlled compression is simply realized
by translations of the walls with the velocity $v_{\mathrm{wall}}$,
so that the packing fraction increases continuously.
%
%%%%%%%%%%%%%%%%%%%%%%%%%%%%%%%%%%%%%%%%%%%%%%%%%%%%%%%%%%%%%%%%%%%%%%%%%%%%%%%%%%%%%%%%%%%%%%%%%%%%

When it reaches a target packing fraction $\phi_k$, 
the translations of the walls are suspended to await the equilibrium.
Usually, the excess stress due to the quickness of the compression
is relaxed by spreading the deformations into the entire system.
The following criteria are examined every $m_{\mathrm{exam}}$ time steps of the simulation:
\begin{itemize}
  \item
  The relative difference between the top and bottom stresses
  is smaller than a threshold value $A $:
  $ |P_{\mathrm{top}} - P_{\mathrm{bot}}|/P  < A $,
  where $P$ is the average, $ P  \equiv (P_{\mathrm{top}} +P_{\mathrm{bot}})/2$.
  \item
  The relative change of the compressive stress 
  for the last $m$ steps
  is smaller than a threshold value $B$:
  $ |P - P'|/ P < B $,
  where $P'$ is the one of $m_{\mathrm{exam}}$ steps before.
  \item
  No bonds were broken for the last $m_{\mathrm{exam}}$ steps.
\end{itemize}
Thus, the compressive stress $P_k$ 
balancing the network of packing fraction $\phi_k$ can be determined.
Once $P_k$ is determined,
the compression continues again for the next target $\phi_{k+1}$.

\subsection{Non real-time simulation for quasi-static simulation}
\label{141932_2Nov12}

In order to reduce finite size effects,
the simulation box needs to involve
a sufficient number of typical structures.
The compressive stress to balance sparse networks is very low,
and the most of particles are slightly stressed.
In this case, the relaxation time is significantly long.
Usually, 
the overdamped equations of motion, 
which are given by dropping the right hand sides of eqs.~\eqref{180838_18Nov12}
due to the negligible inertia of particles,
are used to simulate colloidal suspensions.
However, we have taken a different approach for efficiency.
As restricting our scope in the slow compression,
we may be allowed to optimize some parameters 
for a faster simulation.
In the slow compression limit, 
velocities of particles are also slow.
In this case, 
the contact forces dominate the displacements of particles.
This is why a reduced viscosity $\eta_0'$ still leads to 
the same equilibrium configurations with a shorter time.
The following viscosity gives the critical damped motion 
in terms of the rolling mode.
\begin{equation}
  \eta_0^{\mathrm{(cd)}} = \frac{\sqrt{k_{\mathrm{R}} m}}{3 \pi a }.\label{231133_29Dec12}
\end{equation}
Though it is not simple problem to find the best value,
we used $\eta_0' = 0.1 \eta_0^{\mathrm{(cd)}} $ 
in the equation of motion \eqref{180838_18Nov12}.
This arbitrary selection of the viscosity is justified
only if the compression is considered slow enough.
This point will be checked in \secref{141944_21Nov12}.

\section{Results and discussion}

\subsection{Parameters for contact model}
\label{sec_three-point_bending}

The introduced 2D contact model (see \secref{112400_22Dec12})
is relatively simple,
but it still has 7 parameters to be given.
Traditional experimental measurements
to characterize the particle-particle adhesion
determine the pull-off force~\citep{Larsen_1958},
which corresponds to $F_{\mathrm{Nc}}$ in this work.
In recent years the atomic force microscope became 
a reliable method,
which provides more precise data for 
normal force-displacement relationships~\citep{Hodges_2002}.
However, there are fewer experimental measurements
to determine the parameters of the other modes~\citep{Heim_1999,Ecke_2001}.
The experimental measurement of the three-point bending test
by using optical tweezers
is remarkable~\citep{Pantina_2005,Furst_2007},
which provides parameters for the rolling mode, 
$M_{\mathrm{Rc}}$ and $k_{\mathrm{R}}$ (or $\varphi_{\mathrm{c}}$).
In any case, all parameters for a single system
have never been determined so far.
Thus our knowledge of the contact forces is still limited,
so that we need to set the parameters by hand.
But, we may have a certain level of confidence
by reproducing a similar bending behavior 
with the three-point bending measurements.
Our selected parameters are intended to imitate the observed mechanical behavior:
a small extent of elastic deformation 
and irreversible reorganization due to a critical moment.

%%%%%%%%%%%%%%%%%%%%%%%%%%%%%%%%%%%%%%%%%%%%%%%%%%%%%%%%%%%%%%%%%%%%%%%%%%%%%%%%%%%%%%%%%%%%%%%%%%%%
We have selected two sets of the parameters called bond 1 and bond 2, 
which are shown in \tabref{bond_parameters}.
Bond 1 has the same strengths for the three modes,
while bond 2 is 5 times stronger for the normal and sliding mode than the rolling mode.
Both have the same maximum strains;
elastic deformations give 5\% of the radius.

%%%%%%%%%%%%%%%%%%%%%%%%%%%%%%%%%%%%%%%%%%%%%%%%%%%%%%%%%%%%%%%%%%%%%%%%%%%%%%%%%%%%%%%%%%%%%%%%%%%%

\begin{table}[tbh]
  \caption{
    Parameters for the contact model.
    $C$ is the parameter in eq.~\eqref{eq_f_normal_foo}
    to avoid significant overlap by compression.
  }
  \label{bond_parameters}
  \centering
  \newcolumntype{C}{>{\centering\arraybackslash}X}
  \newcolumntype{R}{>{\raggedright\arraybackslash}X}
  \newcolumntype{L}{>{\raggedleft\arraybackslash}X}
  \begin{tabularx}{12cm}{lCCC}
    \hline 
    & & bond 1 & bond 2 \\
    \hline  
    Normal strength  & $F_{\mathrm{Nc}} / F_0$ & 1 & 5 \\
    Sliding strength & $F_{\mathrm{Sc}}/ F_0$  & 1  & 5\\
    Rolling strength & $M_{\mathrm{Rc}}/ F_0$  & 1  & 1\\
    Normal strain    & $ \delta_{\mathrm{c}}/ a$ & 0.05 & 0.05 \\
    Sliding strain&  $ d_{\mathrm{c}}/a$ & 0.05  & 0.05\\
    Rolling strain&  $ \varphi_{\mathrm{c}}$ & 0.05  & 0.05\\
    Parameter in eq.~\eqref{eq_f_normal_foo} & $C (a^3/F_0)$ & $8 \times 10^4$ & $8 \times 10^4$ \\
    \hline
  \end{tabularx} 
\end{table}

%%%%%%%%%%%%%%%%%%%%%%%%%%%%%%%%%%%%%%%%%%%%%%%%%%%%%%%%%%%%%%%%%%%%%%%%%%%%%%%%%%%%%%%%%%%%%%%%%%%%

\begin{figure}[htb]
  \centering
  \includegraphics[height=144pt]{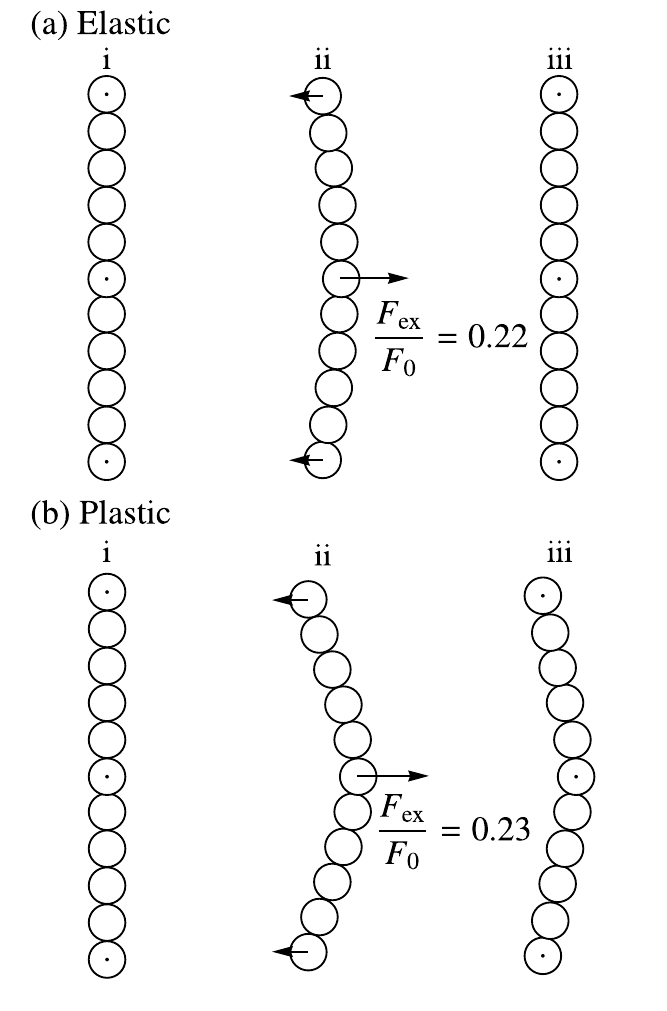}
  \caption{The three-point bending test for a linear aggregate
    consists of the three steps i--iii.
    When a smaller external force ($F_{\mathrm{ex}} = 0.22 F_0$)
    is applied, the elastic deformation is seen (a),
    and when it is slightly larger ($F_{\mathrm{ex}} = 0.23 F_0$),
    the bond breakups cause the plastic deformation (b).
  }
  \label{fig_three-point-bending}
\end{figure}

The simulation for the three-point bending test
consists of the following three steps:
\begin{enumerate}
  \item[i.] 
  Prepare a linear chain under no stress.
  \item[ii.] 
  Apply the forces $(F_{\mathrm{ex}},0,0)$ at the central particle
  and $(-F_{\mathrm{ex}}/2,0,0)$ at both the ends, and wait for equilibrium.
  \item[iii.]
  Stop applying the forces, and wait for equilibrium.
\end{enumerate}
%
%%%%%%%%%%%%%%%%%%%%%%%%%%%%%%%%%%%%%%%%%%%%%%%%%%%%%%%%%%%%%%%%%%%%%%%%%%%%%%%%%%%%%%%%%%%%%%%%%%%%
The three snapshots i--iii in \figref{fig_three-point-bending} (a) and (b)
show the configurations of the respective equilibriums 
of the bond 1 simulation.
Though the data is not shown here,
the results of bond 2 are similar.
The two results were obtained with 
two slightly different forces:
(a) $F_{\mathrm{ex}}/F_0 = 0.22$, (b) $F_{\mathrm{ex}}/F_0 =  0.23$.
If the external force did not reach the threshold,
the initial linear shape remained after the test (see (a)\,iii).
On the other hand,
if the bond stress exceeded the threshold value,
some irreversible rearrangement took place.
As a result, the initial shape was not recovered (see (b)\,iii).
In the latter case,
the bending deformation proceeds 
until the moment acting on the broken bond
becomes smaller than the threshold due to the increase of the bending angle.
This is why the small increment of the applied
forces leads to the visible difference of the deformation
(see ii of (a) and (b) in \figref{fig_three-point-bending}).
The slight bending deformation seen by (a)\,ii
is close to the limit of elastic deformation 
of this cluster.
%

%%%%%%%%%%%%%%%%%%%%%%%%%%%%%%%%%%%%%%%%%%%%%%%%%%%%%%%%%%%%%%%%%%%%%%%%%%%%%%%%%%%%%%%%%%%%%%%%%%%%

Thus, the parameters were chosen
to simulate some brittle behavior of colloidal aggregates;
the local structures remain unless bond breakups take place.
In this test, the rolling mode of the contact model
is responsible for the cluster's deformation.

%%%%%%%%%%%%%%%%%%%%%%%%%%%%%%%%%%%%%%%%%%%%%%%%%%%%%%%%%%%%%%%%%%%%%%%%%%%%%%%%%%%%%%%%%%%%%%%%%%%%

\subsection{Parameters of the compression simulation}
\label{100129_27Dec12}

The initial configurations (\secref{sec_preparation_fractal_gel}) are 
fractal-like networks with a packing fraction 
$\phi_0 \approx 0.12 $,
%$\phi_0 = N \pi / L_x  L_y \approx 0.12 $,
where $(L_x, L_y) = (300a, 500a)$ and $N = 5760$ for bond 1 simulation.
Since the calculation cost of the bond 2 simulation is higher,
smaller system size, $(L_x, L_y) = (200a, 500a)$ and $N=3840$,
is simulated.
The first target packing fraction is set to $\phi_1 = (4/3)\phi_0 \approx 0.16$.
The increment ratio and the number of compression steps 
are $w = 1.1$ and $k' =  17$, respectively.
The compression finally reaches to $\phi_{\mathrm{max}} \approx 0.74$.
The relaxation time depends on the parameters of the bonds.
To account for this, 
the wall velocities are set to $v_{\mathrm{wall}}/v_0 = 5 \times 10^{-4}$
and $v_{\mathrm{wall}}/v_0 = 1 \times 10^{-4}$
for the bond 1 and bond 2 simulations, respectively,
where $v_0 \equiv a \sqrt{k_{\mathrm{R}}/m}$.
In the equilibrium checks (\secref{sec_uniaxial_compression}) 
after every $m_{\mathrm{exam}}=10^4$ time integration steps,
$A = 10^{-2}$ is used for both the simulations,
and $B= 10^{-3}$ and $10^{-4}$ for bond 1 and for bond 2 simulations,
respectively.

\subsection{Uniformity of compression}
\label{141944_21Nov12}

As explained in \secref{141932_2Nov12},
it is useful to introduce artificial parameters
to speed up the compression simulation.
Therefore we started to check that 
our method reproduced the quasi-static compression as expected.
If compression is too fast,
the regions near the wall are easily crushed.
So, we examined the uniformity of the deformations along the compression axis.
The uniformity of compression can be seen via the $y$ dependence 
of the packing fraction $\bar{\phi}(y)$, 
which is defined as the packing fraction in the sliced ranges: 
$ y - \delta L/2 < y' < y + \delta L/2$
(where $\delta L = L_y / 20$ was chosen).
Due to the heterogeneous structure and finite system size,
the uniformity cannot be judged with a single compression simulation.
This is why we took the averages over 10 simulation runs 
with different initial configurations.
%

%%%%%%%%%%%%%%%%%%%%%%%%%%%%%%%%%%%%%%%%%%%%%%%%%%%%%%%%%%%%%%%%%%%%%%%%%%%%%%%%%%%%%%%%%%%%%%%%%%%%

\begin{figure}[tbh]
  \centering
  \includegraphics[height=156pt]{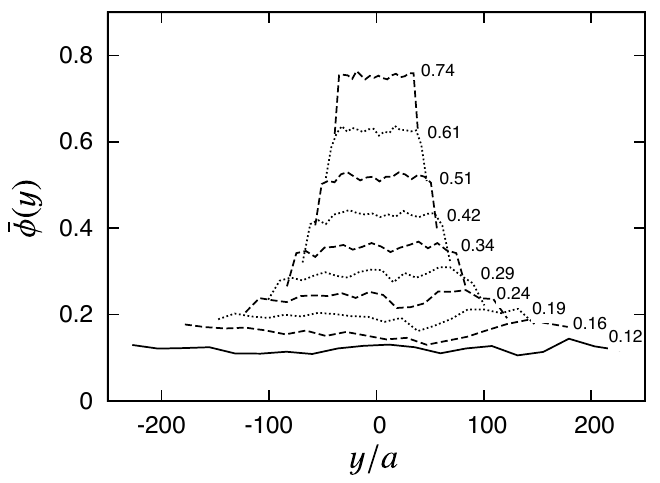}
  \caption{
    The $y$ dependence of packing fractions $\bar{\phi}(y)$
    are shown
    for every second compression 
    step ($\phi_i$, $i=1,3,5\dotsc$).
    The solid line indicates the initial configuration.
  }
  \label{232801_24Oct12}
\end{figure}

\figref{232801_24Oct12} shows the results for the bond 1 simulations,
where every second compression step is plotted.
Although we can see an undesirable drop at the middle
of the first compression ($\phi = 0.16$),
the compression is satisfactorily uniform along $y$-axis after that.
The steep drops at both the edges relate to the integrity of the walls;
particles connecting to the walls are not allowed to move
(see section~\ref{sec_uniaxial_compression}).
Equivalent results were also obtained with the bond 2 simulation
with a slower compaction speed (see \secref{100129_27Dec12}).

\subsection{Data of compressive consolidation}
\label{sec_consolidation}

\begin{figure}[tb]
  \centering
  \includegraphics[height=168pt]{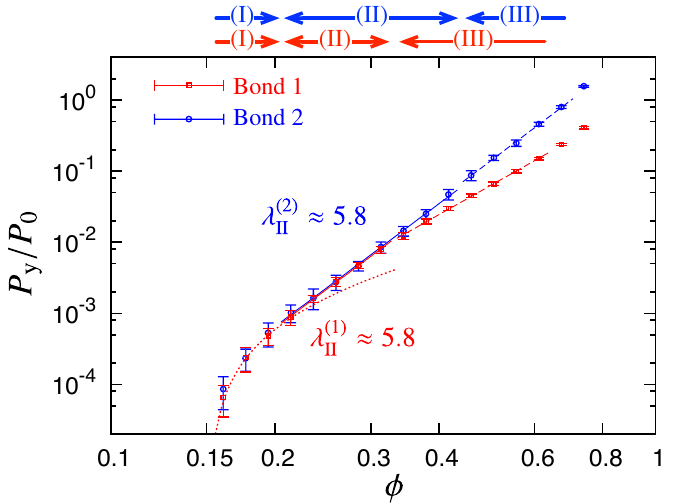} 
  \caption{
    The compressive yield stress $P_{\mathrm{y}} (\phi)$
    is evaluated by the simulations of stepwise compression.
    Its averages and standard deviations taken over 10 runs are shown.
    The red squares represent the result of the bond 1 simulations,
    and the blue circles of the bond 2 simulations.
    The dotted curve shows the fitting function of eq.~\eqref{204301_4Nov12},
    and (solid and dashed) straight lines show the power-law functions \eqref{232118_4Nov12}.
    The arrows indicate
    ranges of the three regimes:
    (I) elastic-dominant regime,
    (II) single-mode plastic regime,
    and (III) multi-mode plastic regime.
    The average exponents $\lambda$ 
    are written on the figure.
  }
  \label{fig_compression_curves}
\end{figure}

Once the system reaches a static equilibrium under a compressive stress $P$,
further compression requires a finite increment: $P \to P + \Delta P$.
Therefore, the balancing stress $P$ can be considered 
as the strength of the gel network,
i.e., the \emph{compressive yield stress} $P_{\mathrm{y}}$~\citep{Buscall_1987}.
A simulation run of the stepwise compression (\secref{141932_2Nov12})
finds a series of $P_{\mathrm{y}}$~\textit{vs.}~$\phi$ relations.
\figref{fig_compression_curves} plots the results of the bond 1 and bond 2 simulations,
where the unit of the stress is $P_{0} =  F_0/a^2$.
The compression from $\phi \approx0.16$ to $\phi \approx 0.74$
shows wide ranges of $P_{\mathrm{y}}$ (about four decades).
These results indicate significantly rapid compressive-strain hardening.

%%%%%%%%%%%%%%%%%%%%%%%%%%%%%%%%%%%%%%%%%%%%%%%%%%%%%%%%%%%%%%%%%%%%%%%%%%%%%%%%%%%%%%%%%%%%%%%%%%%%

Simulation provides various information, which allow us to understand the compression.
Some of data indicate that there are three regimes in the compression process,
which will be seen in the following section.
%
%Let us see them below.

%Simulation provides various data,
%which allow us to understand the compression.
%
%Some of data indicate
%that there are three regimes in the compression process.
%
%Let us see them the below.
%%%%%%%%%%%%%%%%%%%%%%%%%%%%%%%%%%%%%%%%%%%%%%%%%%%%%%%%%%%%%%%%%%%%%%%%%%%%%%%%%%%%%%%%%%%%%%%%%%%%

\paragraph*{Bond creation}

\begin{figure}[tb]
  \centering
  \includegraphics[height=156pt]{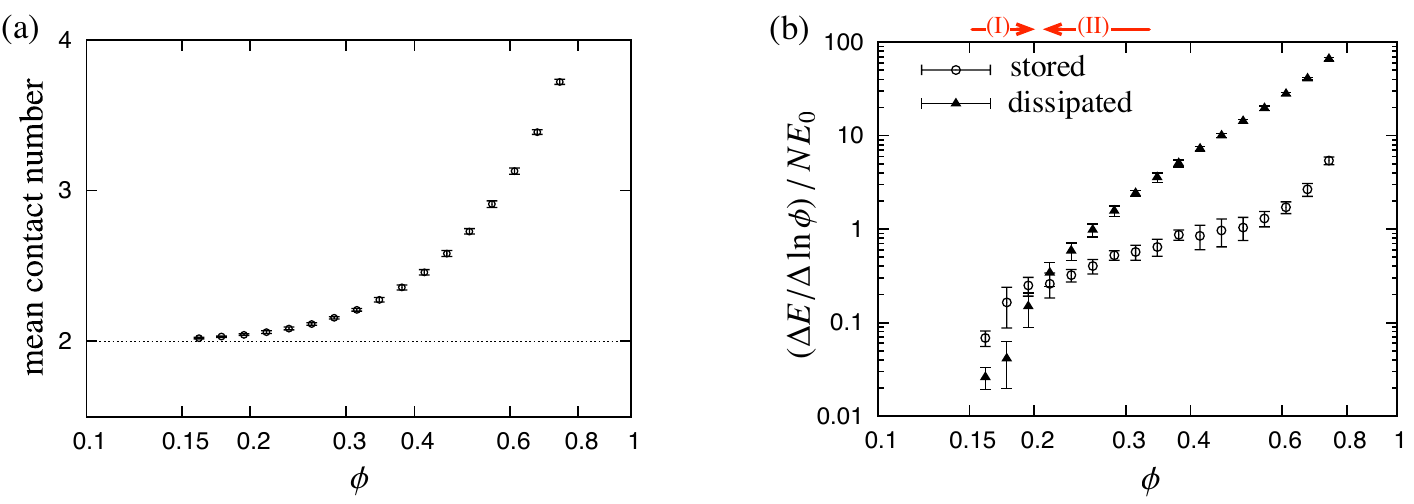} 
  \caption{
    (a)
    The mean contact number per particles
    increases as the compression proceeds.
    The standard deviations are taken over
    the averages of 10 simulations.
    (b)
    The successive increments of 
    mean bond stored and mean dissipated energy rates are shown,
    where $E_0 = k_{\mathrm{B}}a^2 \varphi_{\mathrm{c}}^2/2$.
    The rates are obtained by normalizing
    with the compressive strain $\Delta \log \phi$.
    The averages and standard deviations are taken over
    10 runs of the bond 1 simulation.
  }
  \label{fig_contact_number_energy}
\end{figure}

Bond creation is essential
to understand the irreversible consolidation of the compression.
Due to heterogeneous structure of the networks,
the compressive deformations are not affine.
Consequently, two separated particles may come in contact
and newly bond with each other.
The mean contact number per particle
reflects new bond creations.
The result of bond 1 simulation
is shown in \figref{fig_contact_number_energy}~(a).
The compression continuously changes 
the value from 2 to about 4 with increasing rate,
so we do not see distinct regimes in this.

%%%%%%%%%%%%%%%%%%%%%%%%%%%%%%%%%%%%%%%%%%%%%%%%%%%%%%%%%%%%%%%%%%%%%%%%%%%%%%%%%%%%%%%%%%%%%%%%%%%%

\paragraph*{Stored and dissipated energies}

The bonds have finite strength,
and they can be broken up 
or irreversibly reorganized by certain applied stress.
Such events cause the plastic deformation of the networks,
which contribute the irreversibility.
The translations of the wall by external stress
inject energy into the system.
Part of the energy are stored in contact bonds first
and dissipated as bond ruptures.
We can see this energy flow in the successive increments
of the total stored and dissipated energies
for each compression step~(\figref{fig_contact_number_energy}~(b)).
The total dissipated energy exceeds the total stored energy 
at the early stage, $\phi \approx 0.2$.
Accordingly,
we can identify the boundary of two regimes:
elastic dominant regime (I) and plastic regime (II).
%

%%%%%%%%%%%%%%%%%%%%%%%%%%%%%%%%%%%%%%%%%%%%%%%%%%%%%%%%%%%%%%%%%%%%%%%%%%%%%%%%%%%%%%%%%%%%%%%%%%%%

\paragraph*{Bond rupture}

\begin{figure}[htb]
  \centering
  \includegraphics[height=156pt]{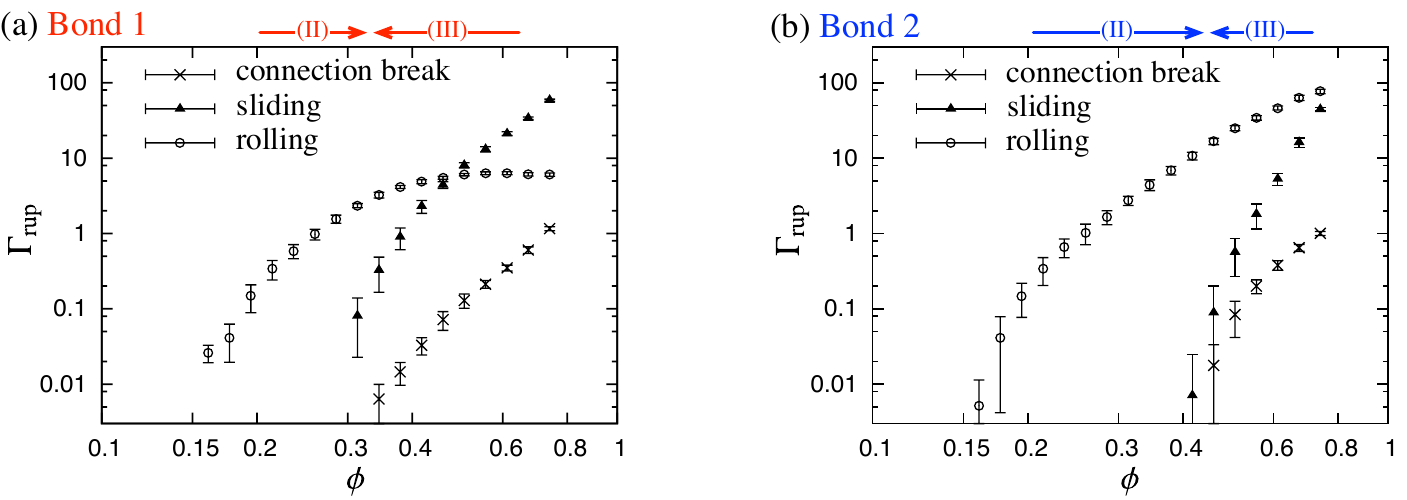}
  \caption{Bond rupture rates $\Gamma_{\mathrm{rup}}$
    are compared for two types of bond: (a) bond 1 and (b) bond 2.
    The bond ruptures are recorded by the causing stress.
    The rates of the separation due to normal forces are plotted by squares,
    the ones of the regeneration due to sliding forces or rolling moments
    by triangles and circles, respectively.
  }
  \label{bond_ruptures}
\end{figure}

The accumulated numbers of the bond ruptures
$N_{\mathrm{rup}}^{(\mathit{mode})}$
were recorded as distinguishing the stress modes causing the breakup,
where the $\mathit{mode}$ indicates 
`connection break', `sliding', or `rolling'.
The rupture rates
$\Gamma_{\mathrm{rup}}^{(\mathit{mode})}$
can be defined as the number of breakups 
normalized with the total particle number and relative compressive strain:
\begin{equation}
  \Gamma_{\mathrm{rup}}^{(\mathit{mode})} \equiv
\frac{1}{N}
  \frac{\mathrm{d} N_{\mathrm{rup}}^{(\mathit{mode})}}{\mathrm{d} \ln \phi}.
\end{equation}
\figref{bond_ruptures} shows
the results of both the bond 1 and bond 2 simulations.
At lower packing fractions, 
only rupture events occur due to rolling.
This indicates that compression of loose networks is achieved 
by bending deformations of particulate chains.
When the compaction reaches a certain point,
further compression requires sliding or separation of contacting particles.
The onset of this mixed-rupture mode 
depends on the parameters of the contact forces.
The connection breaks start between $\phi = 0.31$ and $0.34$
in bond 1 simulations, 
and between $\phi = 0.42$ and $0.46$ in bond 2 simulations.
They split the plastic deformation into two parts:
single-mode plastic regime (II) below
and 
multi-mode plastic regime (III) above.

%%%%%%%%%%%%%%%%%%%%%%%%%%%%%%%%%%%%%%%%%%%%%%%%%%%%%%%%%%%%%%%%%%%%%%%%%%%%%%%%%%%%%%%%%%%%%%%%%%%%

\subsection{Three stages of compression}

The date shown in the previous section
indicate that 
three distinct behaviors 
can be recognized in the compression process.
Let us discuss them one by one here.

\paragraph*{(I) Elastic dominant regime}

\begin{figure}[tbh]
  \centering
  \includegraphics[width=\textwidth]{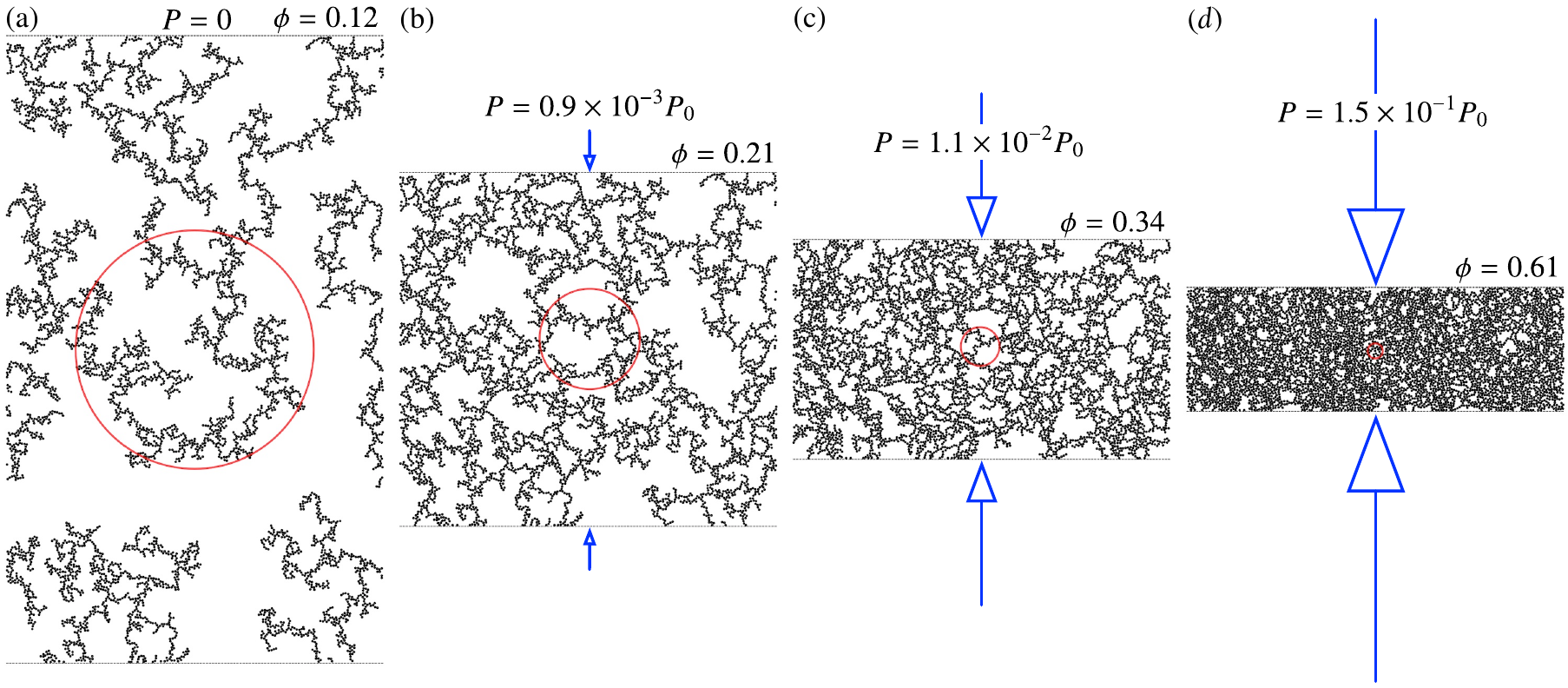}
  \caption{%
    Snapshots of the equilibrium configurations
    under uniaxial compression.
    The initial configuration (a) is a prepared sample in \secref{sec_preparation_fractal_gel},
    whose packing fraction is $\phi \approx 0.12$.
    The configurations (b) and (c)
    show the beginning of 
    the \emph{single-mode plastic regime} ($\phi \approx 0.21$)
    and the \emph{multi-mode plastic regime} ($\phi \approx 0.34$), respectively,
    and (d) shows $\phi \approx 0.61$.
    The red circles express the correlation lengths;
    their diameters are $4 \xi^{(0.5)}$,
    which will be evaluated 
    in the later section (\secref{sec_structure_compression}).
  }
  \label{fig_snapshots}
\end{figure}

There is no percolation path in the 10 initial configurations of $\phi_0 \approx 0.12$
(see \figref{fig_snapshots} (a)).
Hence, no compressive stress is required 
to compress them in the slow compression limit.
After some compaction, 
the first spanning path appears, 
and the network begins to resist the compressive stress.
At the gel point,
a single-linking chain (percolation path)
resists the external stress
between top and bottom.
This system-spanning cluster is deformed as a whole,
so that strains of single bonds can be very small.
Therefore, the dominant response to the external stress 
is elastic deformation;
the work by compression 
is stored in the contact bonds~(see \figref{fig_contact_number_energy}~(b)).
%
%%%%%%%%%%%%%%%%%%%%%%%%%%%%%%%%%%%%%%%%%%%%%%%%%%%%%%%%%%%%%%%%%%%%%%%%%%%%%%%%%%%%%%%%%%%%%%%%%%%%

In this range, the power-law form expanding at the gel point $\phi_{\mathrm{g}}$
is suitable to capture the $P_{\mathrm{y}}$-$\phi$ curve~\citep{de-Gennes_1976}:
\begin{equation}
  P_{\mathrm{y}}/P_0 =
  C' (\phi-\phi_{\mathrm{g}})^{\lambda'},
  \quad
  \text{for $\phi > \phi_{\mathrm{g}}$}.
  \label{204301_4Nov12}
\end{equation}
%
%This function allows us 
%to estimate the gel point $\phi_{\mathrm{g}}$ 
%by extrapolation. 
%
In \figref{fig_compression_curves},
the dotted curve shows the fitting result
with the function \eqref{204301_4Nov12}
for the bond 1 simulation data of $\phi_{\mathrm{g}} < \phi < 0.2$.
The fitting gives the extrapolated gel point as $\phi_{\mathrm{g}} \approx 0.15 $.
The evaluated exponent $\lambda' \approx 1.5 (\neq 1)$ 
represents some nonlinearity.
The compression induces some new contacts,
so that the numbers of percolation paths are increased.
The rise of resistance due to such increasing paths
can be a plausible interpretation of the nonlinearity 
in this elastic-dominant regime.
%

%%%%%%%%%%%%%%%%%%%%%%%%%%%%%%%%%%%%%%%%%%%%%%%%%%%%%%%%%%%%%%%%%%%%%%%%%%%%%%%%%%%%%%%%%%%%%%%%%%%%

It must be noted that
our simulation results in this range
might have large statistical fluctuations.
The typical length scale of the structure is
comparable to the system size
(compare the drawn circle and the box size 
in \figref{fig_snapshots}~(a)),
and consequently only a few paths enter the averaging.
Therefore,
the gel point $\phi_{\mathrm{g}}$ obtained in the above fitting 
has only limited significance.

%
%This is seen in the error bars in \figref{fig_compression_curves}.
%

%%%%%%%%%%%%%%%%%%%%%%%%%%%%%%%%%%%%%%%%%%%%%%%%%%%%%%%%%%%%%%%%%%%%%%%%%%%%%%%%%%%%%%%%%%%%%%%%%%%%

\paragraph*{(II) Single-mode plastic regime}

By considering 
two data (\figref{fig_contact_number_energy}~(b) and \figref{bond_ruptures}),
we identified the single-mode plastic regime
in the ranges: $0.2 < \phi < 0.31 $ for bond 1 simulations
and $0.2 < \phi < 0.42 $ for bond 2 simulations.

%%%%%%%%%%%%%%%%%%%%%%%%%%%%%%%%%%%%%%%%%%%%%%%%%%%%%%%%%%%%%%%%%%%%%%%%%%%%%%%%%%%%%%%%%%%%%%%%%%%%

This $\phi$-dependence of the consolidation behavior
has been discussed with a scaling concept~\citep{Buscall_1988,Shih_1990a}.
The power-law relationships of $P_{\mathrm{y}}$ and $\phi$ 
have been also observed in experiments~\citep{Buscall_1987a,Madeline_2007,Parneix_2009}:
\begin{equation}
  P_{\mathrm{y}}(\phi)/P_0 =  C \phi^{\lambda}. \label{232118_4Nov12}
\end{equation}
A similar power-law behavior
can be seen in our simulation results (\figref{fig_compression_curves}).
The least squares fitting 
gives the almost same exponents:
$\lambda_{\mathrm{II}}^{\mathrm{(1)}} \approx 5.8 \pm 0.5$ and 
$\lambda_{\mathrm{II}}^{\mathrm{(2)}} \approx 5.8 \pm 0.4$,
for the bond 1 and bond 2 results, respectively.
%

%%%%%%%%%%%%%%%%%%%%%%%%%%%%%%%%%%%%%%%%%%%%%%%%%%%%%%%%%%%%%%%%%%%%%%%%%%%%%%%%%%%%%%%%%%%%%%%%%%%%

Let us look more closely at this regime.
\figref{fig_snapshots} (b)
shows the configuration of particles
at the beginning of this range ($\phi \approx 0.21$)
in a bond 1 simulation.
At the gel point, a single linking path spanned the entire box.
In contrast, multi-linking structures span the system here.
Since the maximum bending moment of a random-walk chain
is proportional to the inverse of the chain size,
the supportable loads of larger loops are smaller.
The variety of the mesh sizes
indicates a heterogeneity of the local strengths.
Though large empty spaces remain,
the relatively dense meshes fill the space from the bottom to the top
(\figref{fig_snapshots}~(b)).
Further compression requires destroying a part of them.
As shown in \figref{bond_ruptures},
the rolling-mode ruptures are the only rupture events 
seen in this range.
Therefore, the size of the responsible mesh structures
determines the compressive yield stress $P_{\mathrm{y}}$.

%%%%%%%%%%%%%%%%%%%%%%%%%%%%%%%%%%%%%%%%%%%%%%%%%%%%%%%%%%%%%%%%%%%%%%%%%%%%%%%%%%%%%%%%%%%%%%%%%%%%

Thus, in the single-mode plastic regime,
the length scale is essential;
the compression breaks larger and weaker structures 
into smaller and more robust structures.
Fractal model was employed
to give the length scale from 
the packing fraction~\citep{Brown_1986,Shih_1990a}
(see \secref{sec_structure_compression}).
According to the nature of a fractal,
a portion of a fractal cluster is also a small fractal cluster.
If the compression of fractal networks
causes only fragmentation of fractal clusters 
into smaller pieces,
it can be expected that
the correlation length is decreased 
while keeping the same fractal dimension.
This optimistic expectation leads us to employ
the fractal model for the compression problem.
We will check this point in the next section.

\paragraph*{(III) Multi-mode plastic regime}

%%%%%%%%%%%%%%%%%%%%%%%%%%%%%%%%%%%%%%%%%%%%%%%%%%%%%%%%%%%%%%%%%%%%%%%%%%%%%%%%%%%%%%%%%%%%%%%%%%%

The onset of connection breaks indicates that
the compression enters a new regime.
This transition point depends on the parameters of the bonds.
We consider the data above $\phi \approx 0.34$ 
and $\phi \approx 0.46$ in bond 1 and 2 simulations
as the multi-mode plastic regime.
In these ranges,
the compression curves look to follow a power law.
However, 
the evaluated exponents are totally different with each other:
$\lambda_{\mathrm{III}}^{\mathrm{(1)}} \approx 4.3 \pm 0.2$ 
and 
$\lambda_{\mathrm{III}}^{\mathrm{(2)}} \approx 5.8 \pm 0.4$.
It is by chance that
the exponent of bond 2 simulations 
is similar with the one in the single-mode plastic regime.
The exponent in this regime
is easily affected by the parameters of the contact model,
that implies absence of a simple scaling law.

%%%%%%%%%%%%%%%%%%%%%%%%%%%%%%%%%%%%%%%%%%%%%%%%%%%%%%%%%%%%%%%%%%%%%%%%%%%%%%%%%%%%%%%%%%%%%%%%%%%%

\figref{fig_snapshots} (c) depicts 
the beginning of this regime
of a bond 1 simulation ($\phi \approx 0.34$).
There are still large loops remaining,
but they look like voids due to the denser surroundings.
Here, we can easily find dense meshes 
and trimers (a trimer can be considered
as the building block of a close packing).
If they are next to each other,
the entire object can be considered as a lump.
The same scenario of the \emph{single-mode plastic regime} seems to work;
denser mesh domains are piled up over the entire system.
However, due to the finite size of particles,
the mechanical response of denser meshes is not 
the same as looser ones.
The crush of them involves
sliding ruptures and connection breaks,
which are seen in \figref{bond_ruptures}.
Since the criteria of destruction includes the all three modes,
the compression curve depends on the details of the contact force.
%

%%%%%%%%%%%%%%%%%%%%%%%%%%%%%%%%%%%%%%%%%%%%%%%%%%%%%%%%%%%%%%%%%%%%%%%%%%%%%%%%%%%%%%%%%%%%%%%%%%%%

This regime terminates due to the close packing,
where 
rigidity of particles prevents reorganizations of particles.
Since our simplified contact model 
is not accurate for that,
we cannot specify the upper end of the range.

\subsection{Fractal correlation under compression}
\label{sec_structure_compression}

During compression, separated particles can newly 
come into contact.
Consequently, mesh size decreases as $\phi$ is increased by compression.
If we employ the fractal model~\citep{Brown_1986,Shih_1990a}
to explain the $\phi$-dependence of $P_{\mathrm{y}}$,
the structure change needs to follow a special manner;
the fractal dimension $D_{\mathrm{f}}$ 
characterizing the local structures
does not depend on the packing fraction $\phi$,
and the correlation length $\xi$
is a decreasing function of the packing fraction $\phi$,
$\xi \sim \phi^{-1/(d-D_{\mathrm{f}})}$.

%%%%%%%%%%%%%%%%%%%%%%%%%%%%%%%%%%%%%%%%%%%%%%%%%%%%%%%%%%%%%%%%%%%%%%%%%%%%%%%%%%%%%%%%%%%%%%%%%%%%

The fractal dimension $D_{\mathrm{f}}$ %and correlation length $\xi$
can be estimated from the density-density correlation function $C(r)$
(see \appendixname~\ref{appendix_density-correlation}).
\figref{fig_ddc} shows averages of the correlation functions $C(r)$ of 
every second compression step, $\phi_i$ ($i=1,3,5, \dotsc$),
of the bond 1 simulation results.
The results of the bond 2 simulations were almost the same. 
The correlation of the density is mostly decreasing functions of 
the two-point distance $r$.
The correlation of the configuration before compression is the largest
and reaches the farthest distance (solid line).
The compression reduces this correlation in a continuous manner.

\begin{figure}[htb]
  \centering
  \includegraphics[height=156pt]{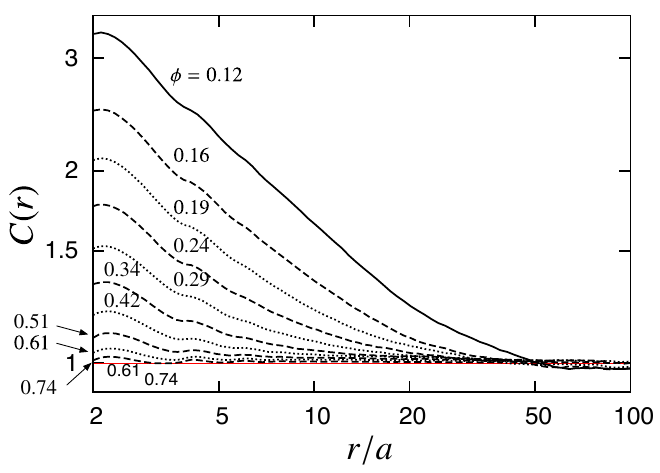}
  \caption{
    The density-density correlation functions $C(r)$
    of compressed networks are shown.
    The solid line shows the initial configuration $\phi_0$,
    and the dashed or dotted lines
    the every second compression step $\phi_i$ ($i=1,3,5,\dotsc$).
    They are averaged values over 10 runs of simulations.
  }
  \label{fig_ddc}
\end{figure}

%%%%%%%%%%%%%%%%%%%%%%%%%%%%%%%%%%%%%%%%%%%%%%%%%%%%%%%%%%%%%%%%%%%%%%%%%%%%%%%%%%%%%%%%%%%%%%%%%%%%

\begin{figure}[htb]
  \centering
  \includegraphics[height=156pt]{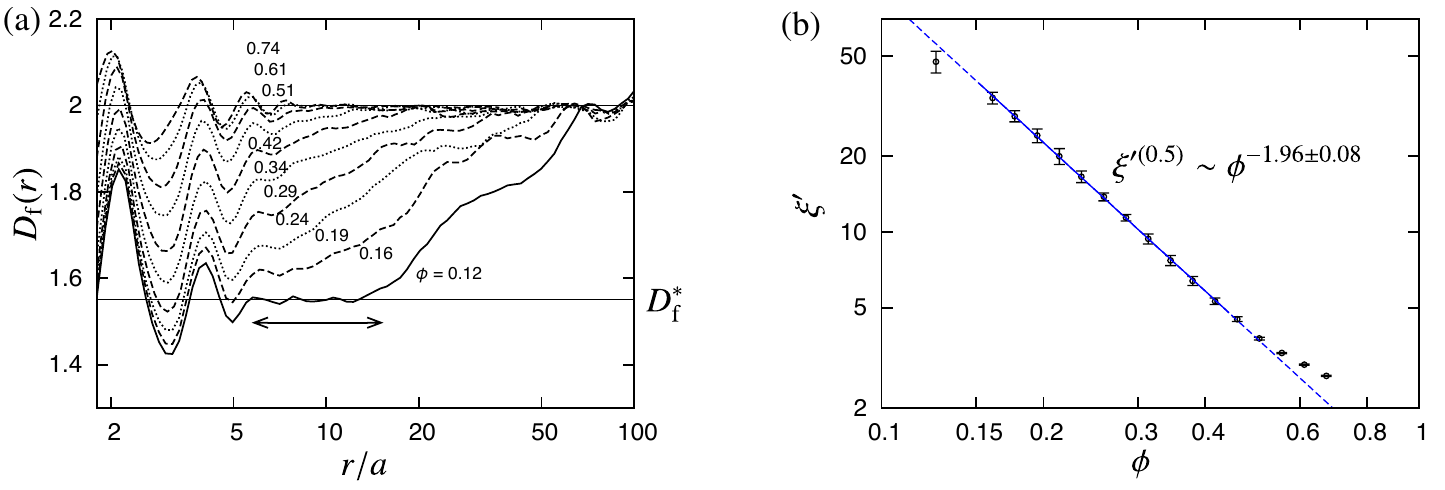}
  \caption{
    (a) The fractal dimension profiles $D_{\mathrm{f}}(r)$
    evaluated by eq.~\eqref{r-dep_fractal_dimension} are plotted.
    The solid line shows the initial configuration $\phi_0$,
    and the dashed or dotted solid lines 
    the every second compression step $\phi_i$ ($i=1,3,5,\dotsc$).
    The numerical values indicate the packing fractions.
    (b) The $\phi$-dependence of 
    the correlation length $\xi^{(0.5)}$ (see the definition in the text).
  }
  \label{143513_26Nov12}
\end{figure}

The fractal correlation appears
as a straight slope in the log-log plot of $C(r)$,
i.e., $C(r) \sim r^{-(d - D_{\mathrm{f}})}$.
Though the straight lines can be recognized
in the initial configuration,
the compressed networks show slightly curved lines.
In order to make them easier to see,
we introduce the fractal dimension profile $D_{\mathrm{f}}(r)$
from the local slope of the correlation function at each distance $r$:
\begin{equation}
  D_{\mathrm{f}}(r) \equiv d + \frac{\mathrm{d} \ln C(r)}{\mathrm{d} \ln r}.
  \label{r-dep_fractal_dimension}
\end{equation} 
The results are shown in \figref{143513_26Nov12}~(a).
As mentioned in \secref{sec_preparation_fractal_gel},
the initial configurations (solid line)
consist of three parts:
a fractal plateau ($ 6< r/a < 15$) 
indicating $D_{\mathrm{f}}^{\ast} \approx 1.55$,
crossover range ($15< r/a < 70$),
and 2D-like uniform network ($r/a > 70$).
Thus, though the recognizable fractal plateau is limited,
%we can confirm that 
the $D_{\mathrm{f}}(r)$ profiles
of the initial configurations
agree with the expected behavior as fractal networks.

%%%%%%%%%%%%%%%%%%%%%%%%%%%%%%%%%%%%%%%%%%%%%%%%%%%%%%%%%%%%%%%%%%%%%%%%%%%%%%%%%%%%%%%%%%%%%%%%%%%%

Once the networks are compressed,
the fractal correlations are affected immediately.
Indeed,  
the first compression step of our simulation ($\phi \approx 0.16$)
shows the entire change of $D_{\mathrm{f}}(r)$:
a narrower plateau indicating a rise of the fractal dimension,
and shifts of the crossover and uniform ranges to shorter lengths.
These results suggest that
not only long-distance correlations, 
but short-distance correlations can be affected by the compressive deformation
at the early stage.
This tendency continues for the further compression.

%%%%%%%%%%%%%%%%%%%%%%%%%%%%%%%%%%%%%%%%%%%%%%%%%%%%%%%%%%%%%%%%%%%%%%%%%%%%%%%%%%%%%%%%%%%%%%%%%%%%

Though the profiles $D_{\mathrm{f}}(r)$ for the compressed networks
indicate that mono-fractal assumption does not work,
it may be worth determining
the $\phi$ dependence of the correlation length $\xi$.
Since the slopes of $C(r)$ in \figref{fig_ddc} 
are not constant,
it is ambiguous to determine $\xi$ from them.
In our purpose,
we may consider the correlation length 
as the length scale where the network looks uniform.
The typical length scale for the uniformity
can be determined by evaluating
fluctuation of local densities~\citep{Masschaele_2009}.
The local densities are defined 
by boxes of size $l$, which are denoted by $\rho_i(l)$.
If the size of box is larger than the correlation length $\xi$,
the local densities show little fluctuation; the system looks uniform.
Thus, the standard deviation of the local densities
normalized by the global density,
$ \Delta \rho(l) /\langle \rho \rangle$, measures the uniformity.
Due to the finite system size,
we compare here the lengths showing a common finite fluctuation, $0.5$.
The actual values of the correlation lengths $\xi$
should be larger than the evaluated lengths $\xi'^{(0.5)}$,
so a common arbitrary factor may be given to adjust.
\figref{143513_26Nov12}~(b) shows
the result of the lengths $\xi'^{(0.5)}$ satisfying 
$ \Delta \rho(\xi'^{(0.5)}) /\langle \rho \rangle = 0.5$.
We find a power-law relation 
between $\xi$ and $\phi$ with an exponent $-1.96 \pm 0.08$.
The obtained exponent is a little bit smaller than the fractal assumption: 
$-1/(d-D_{\mathrm{f}}^{\ast}) \approx -2.2$.
%though we saw that the fractal dimension $D_{\mathrm{f}}^{\ast}$
%is valid only for the initial configuration.
%
The evaluated correlation lengths, $4\xi'^{(0.5)}$,
are also shown as the diameter of the circles drawn in \figref{fig_snapshots}.

\section{Conclusion}

We focused on investigating 
compression of particle gels.
Inspired by scaling theory for polymer gels,
the fractal model has been introduced
to explain the packing fraction dependence
of particle gels' rheology~%
\citep{Brown_1986,Shih_1990a},
and referred for the compression process~\citep{Buscall_1988}.
The model assumes that 
the structure of a particle gel is fractal network.
However,
the simple question whether a particle gel under compression
is fractal network or not,
has left unanswered.
We have aimed to answer this question 
by introducing a particle-scale simulation model 
in the limit of strongly aggregated particle gels.
In this case, the cohesive force among contacting particles 
is assumed to be so strong 
that Brownian forces do not affect
the structural evolution of gels.
%no free rolling is allowed and  motion is negligible.
%

%%%%%%%%%%%%%%%%%%%%%%%%%%%%%%%%%%%%%%%%%%%%%%%%%%%%%%%%%%%%%%%%%%%%%%%%%%%%%%%%%%%%%%%%%%%%%%%%%%%%

Our simulation reproduced
the reasonable consolidation behaviors
as $\phi$ \textit{vs}.~$P_{\mathrm{y}}$ relationships
in the case of quasi-static compression.
By monitoring various particle-scale information,
i.e., 
mean contact number,
stored/dissipated energy,
and number of bond ruptures,
we identified the three distinct regimes:
(I) elastic dominant regime,
(II) single-mode plastic regime,
and (III) multi-mode plastic regime.
At the gel point, there are many ``dead end'' branches,
where the particles in such branches do not contribute
to the strength of the network $P_{\mathrm{y}}$.
Compression makes separated particles come into contact,
so that the mesh size becomes smaller and smaller.
This explanation is already known and qualitatively true for the most part.
Such shrinkage of the typical length scale
causes a transition of the local mechanical response
from (II) to (III) in the plastic regimes.
While the length scale is large enough,
the rolling mode is the responsible for the deformation.
At a certain length scale,
the sliding and normal modes become more important for 
further compression.
Thus, we confirmed that
a single regime could never cover the wide range of packing fractions
in spite of subtle appearance 
in the $\phi$ \textit{vs}.~$P_{\mathrm{y}}$ relationships.
The simulation results also provided an insight that
heterogeneity in the local mesh sizes
plays some roles in the way of reorganization under compression
and overall network strength.
Mechanical correlations (connectivity)
determine which parts of the network are destroyed under an applied stress.
%robust
If a percolation path consisting of robust domains is built,
they support a large fraction of the applied stress.
In this case, the weakest domain in the path 
is most probable to be destroyed.
This restructuring yields a smaller and robuster domain,
which results in the reinforcement of the percolation path.
However, such mechanism is subtle 
because compression is not continuing deformation
and tends to reduce the heterogeneity.

%
%%%%%%%%%%%%%%%%%%%%%%%%%%%%%%%%%%%%%%%%%%%%%%%%%%%%%%%%%%%%%%%%%%%%%%%%%%%%%%%%%%%%%%%%%%%%%%%%%%%%

In order to see whether the fractal assumption remains or not 
in particle gels under compression,
we evaluated 
the density-density correlation $C(r)$
for the simulation results.
The evaluated fractal dimension profile $D_{\mathrm{f}}(r)$
says that compression affects
both the local structure and correlation length
in the early stage.
Thus, our results are contradictory to 
the application of the fractal model to compression processes.
However, 
the fractal dimension profiles $D_{\mathrm{f}}(r)$
show a systematic change
in the \emph{single-mode plastic regime}.
This result seems to indicate required ingredients 
to upgrade the fractal model for compression processes.

%%%%%%%%%%%%%%%%%%%%%%%%%%%%%%%%%%%%%%%%%%%%%%%%%%%%%%%%%%%%%%%%%%%%%%%%%%%%%%%%%%%%%%%%%%%%%%%%%%%%

The simulation model of this work is simple;
we neglect various factors in real particle gels
to make clear the focus of study and 
to optimize the simulation speed
in order to achieve large system sizes.
For example, we employed a simple contact model,
which depends on only configuration of particles
and has no history dependence.
However, cohesion between particles in real system
may be more complex,
e.g.
the bonding or breakup may require certain time periods.
In order to deal with such systems,
our quasi-static simulation framework needs to be extended.
%to introduce rete dependence.
%
This extension is relevant for the future work
because there are many experimental works
investigating on the time/rate-depending aspects of colloidal gels 
and yield stress fluids~%
\citep{Manley_2005,Bonn_2009,Brambilla_2011}.

\begin{acknowledgments}
The authors would like to thank Prof. Morton Denn
for valuable suggestions on the manuscript,
and RS would like to express his gratitude to 
Prof. Richard Buscall
for fruitful discussions, valuable suggestions,
and warm encouragement.
\end{acknowledgments}

\appendix

\section{Density-density correlation function}

\label{appendix_density-correlation}

In order to study colloidal gels,
we need to have some tools to characterize random structured systems.
The density-density correlation function $C(d\bm{r})$ is used for this purpose.
The local density function $\rho(\bm{r})$ is defined as follows:
\begin{equation}
  \rho (\bm{r}) = 
  \begin{cases}
    1, & \text{$\bm{r}$ is in the solid phase}, \\
    0, & \text{$\bm{r}$ is in the liquid phase}.
  \end{cases}
\end{equation}
By taking the average over the entire space of the system,
the density-density correlation function $C(d\bm{r})$ 
is defined as follows:
\begin{equation}
  C(d \bm{r}) 
  \equiv
  \frac{\bigl\langle
      \rho (\bm{r})  \rho (\bm{r}+d\bm{r}) 
      \bigr\rangle}{
    \left\langle \rho (\bm{r})
      \right\rangle^2
  }
\end{equation}
When the system is isotropic,
the correlation is given in terms of the distance, $r = |d\bm{r}|$,
which can be obtained by taking averages over all directions:
\begin{equation}
  C(r)  \equiv 
  \bigl\langle
    C(d \bm{r}) 
    \bigr\rangle_{r = |d\bm{r}|}
\end{equation}
The correlation function $C(r)$
can be regarded as a conditional probability
to find particles.
Since it is normalized by the entire probability,
the value 1 means no correlation.
%

%%%%%%%%%%%%%%%%%%%%%%%%%%%%%%%%%%%%%%%%%%%%%%%%%%%%%%%%%%%%%%%%%%%%%%%%%%%%%%%%%%%%%%%%%%%%%%%%%%%%

It is known that strongly aggregated gels 
have structures involving fractal correlations.
The local structures are equivalent to fractal flocs.
For fractal flocs,
the relation between the number of particles $N$ 
and the radius of gyration $R_{\mathrm{g}}$ 
follows a power-law:
\begin{equation}
  N \sim R_{\mathrm{g}}^{D_{\mathrm{f}}}.
\end{equation}
The exponent $D_{\mathrm{f}}$ is called fractal dimension.
Therefore, the average packing-fraction profile of the clusters
is given as follows:
\begin{equation}
  \phi (r) \sim r^{-(d-D_{\mathrm{f}})},\label{002007_3Jan13}
\end{equation}
where $r$ is the distance from the center of mass.
Since fractal gels can be considered as cluster-filling networks,
the local structure should coincide with this density profile \eqref{002007_3Jan13}.
Consequently, the density-density correlation function $C(r)$
has the same functional form: $C(r) \sim \phi(r)$.
In a fractal gel, the correlation reaches a finite distance, 
which is called the correlation length $\xi$.
For $r >\xi $, the correlation does not remain:
\begin{equation}
  C(r)
  \sim 
  \begin{cases}
    r^{-(d-D_{\mathrm{f}})}, &\text{for $r < \xi$.} \\
    1 ,   &\text{for $r > \xi$.} 
  \end{cases}
\end{equation}
If the local densities are evaluated in boxes of size $\xi$,
the system should look uniform.

%\bibliography{/Users/seto/Dropbox/Papers/rse}

%merlin.mbs aipauth4-1.bst 2010-07-25 4.21a (PWD, AO, DPC) hacked
%Control: author (9) reversed initials

\end{document}